\documentclass[prl,twocolumn,twoside,preprintnumbers,superscriptaddress,nofootinbib,showpacs,showkeys]{revtex4}
\usepackage{amsmath,slashed}
\usepackage{graphicx,graphics}
\usepackage{dcolumn}
\usepackage{bm}
\usepackage[hyperfootnotes=false]{hyperref}
\usepackage{xspace}

\newcommand{\Fpi}{F_\pi}
\newcommand{\mpi}{M_{\pi}}
\newcommand{\mpii}{M_{\pi^0}}
\newcommand{\ga}{g_A}

\newcommand{\muu}{m_u}
\newcommand{\md}{m_d}

\newcommand{\mN}{m_N}
\newcommand{\sm}{s_\text{m}}
\newcommand{\MeV}{\,\text{MeV}}
\newcommand{\GeV}{\,\text{GeV}}
\newcommand{\beq}{\begin{equation}}
\newcommand{\eeq}{\end{equation}}
\newcommand{\Amp}{\mathcal{A}}
\renewcommand{\Re}{\text{Re}\,}

\begin{document}

\preprint{INT-PUB-15-026}
\title{High-precision determination of the pion--nucleon $\boldsymbol{\sigma}$-term from Roy--Steiner equations}

\author{Martin Hoferichter}
\affiliation{Institut f\"ur Kernphysik, Technische Universit\"at Darmstadt, D--64289 Darmstadt, Germany}
\affiliation{ExtreMe Matter Institute EMMI, GSI Helmholtzzentrum f\"ur Schwerionenforschung GmbH, D--64291 Darmstadt, Germany}
\affiliation{Albert Einstein Center for Fundamental Physics, Institute for Theoretical Physics,
University of Bern, Sidlerstrasse 5, CH--3012 Bern, Switzerland}
\affiliation{Institute for Nuclear Theory, University of Washington, Seattle, WA 98195-1550, USA}
\author{Jacobo Ruiz de Elvira}
\author{Bastian Kubis}
\affiliation{Helmholtz--Institut f\"ur Strahlen- und Kernphysik (Theorie) and\\
   Bethe Center for Theoretical Physics, Universit\"at Bonn, D--53115 Bonn, Germany}
\author{Ulf-G.\ Mei{\ss}ner}
\affiliation{Helmholtz--Institut f\"ur Strahlen- und Kernphysik (Theorie) and\\
   Bethe Center for Theoretical Physics, Universit\"at Bonn, D--53115 Bonn, Germany}
\affiliation{Institut f\"ur Kernphysik, Institute for Advanced Simulation, 
   J\"ulich Center for Hadron Physics, JARA-HPC, and JARA-FAME,  Forschungszentrum J\"ulich, D--52425  J\"ulich, Germany}

\begin{abstract}
We present a determination of the pion--nucleon ($\pi N$) $\sigma$-term $\sigma_{\pi N}$ based on the Cheng--Dashen low-energy theorem (LET), taking advantage of
the recent high-precision data from pionic atoms to pin down the $\pi N$ scattering lengths as well as of constraints from analyticity, unitarity, and crossing symmetry in the form of Roy--Steiner equations to perform the extrapolation to the Cheng--Dashen point in a reliable manner. 
With isospin-violating corrections included both in the scattering lengths and the LET, we obtain
$\sigma_{\pi N}=(59.1\pm 1.9\pm 3.0)\MeV = (59.1\pm 3.5)\MeV$, where the first error refers to uncertainties in the $\pi N$ amplitude and the second to the LET. Consequences for the scalar nucleon couplings relevant for the direct detection of dark matter are discussed.
\end{abstract}


\maketitle

\section{Introduction}

The $\pi N$ $\sigma$-term measures the amount of the nucleon mass that is generated by the two lightest quarks.
Since the dominant contribution originates from the energy content of the gluon field, due to the 
trace anomaly of the QCD energy-momentum tensor, the nucleon mass would only change moderately if the quark masses were turned off. 
Thus, $\sigma_{\pi N}$ encodes information on the explicit breaking of chiral symmetry and constitutes one of the fundamental low-energy
parameters of QCD. In recent years, a precise determination of the $\sigma$-term has become increasingly urgent, given its relation to
the scalar couplings of the nucleon that are prerequisite for a consistent interpretation of direct-detection dark matter searches~\cite{Bottino:1999ei,Ellis:2008hf,Crivellin:2013ipa}.

Traditionally, information on $\sigma_{\pi N}$ has been inferred from $\pi N$ scattering by means of the Cheng--Dashen LET~\cite{Cheng:1970mx,Brown:1971pn} that relates the Born-term-subtracted isoscalar amplitude $\bar D^+$ at the Cheng--Dashen point $s=u=\mN^2$, $t=2\mpi^2$, to the scalar form factor of
the nucleon $\sigma(t)$ evaluated at $t=2\mpi^2$ (precise definitions below). The application of the LET thus requires two main ingredients: the analytic continuation of
the isoscalar $\pi N$ amplitude into the unphysical region, and the correction due to the finite momentum transfer in $\sigma(2\mpi^2)$. 
The first task has been addressed by extrapolating partial-wave analyses (PWAs) from the physical region to the Cheng--Dashen point by means of dispersion relations~\cite{Koch:1980ay,Koch:1982pu,Hoehler}, in particular, in~\cite{Gasser:1988jt,Gasser:1990ce} a formalism was developed to express 
the result of the extrapolation in terms of threshold parameters for $\pi N$ scattering. 
Similarly, the scalar form factor requires a dispersive reconstruction to account for the strong $\pi\pi$ rescattering in the isospin-$0$ $S$-wave~\cite{Gasser:1990ap}. 
Based on the PWA from~\cite{Koch:1980ay,Hoehler}, a value $\sigma_{\pi N}\sim 45\MeV$ was inferred in~\cite{Gasser:1990ce}. This result was later challenged by a new PWA~\cite{Pavan:2001wz}, leading to a much larger value of $\sigma_{\pi N}=(64\pm 8)\MeV$, although based on the same formalism. In fact, the discrepancy could be traced back, to about equal parts, to different input for the isoscalar $\pi N$ scattering length, the $\pi N$ coupling constant, and $\pi N$ partial waves for the evaluation of the dispersive integrals. 

A second strategy that has been pursued relies on Chiral Perturbation Theory (ChPT) to perform the extrapolation to the Cheng--Dashen point. 
However, to determine low-energy constants still input for the $\pi N$ phase shifts is required, so that the outcome of the ChPT analyses 
tends to support the value of $\sigma_{\pi N}$ corresponding to the PWA used as input~\cite{Fettes:2000xg,Alarcon:2011zs}. Moreover, it has been questioned whether the chiral representation is at all accurate enough to permit a reliable extrapolation to the Cheng--Dashen point~\cite{Becher:2001hv}. For a detailed comparison to results obtained in lattice QCD,
 we refer to~\cite{inprep}. 

In this Letter, we combine two new sources of information on $\pi N$ scattering that have become available over the last years. First, the measurement of level shifts and decay widths in pionic atoms~\cite{Gotta:2008zza,Strauch:2010vu,Hennebach:2014lsa} has led to a precision determination of the $\pi N$ scattering lengths~\cite{Baru:2010xn,Baru:2011bw}. Second, a system of Roy--Steiner (RS) equations has been developed~\cite{Ditsche:2012fv} that combines general constraints on the $\pi N$ scattering amplitude imposed by analyticity, unitarity, and crossing symmetry. The construction proceeds similarly to Roy equations for $\pi\pi$ scattering~\cite{Roy:1971tc}, where the solution for the low-energy phase shifts can be parameterized in terms of the $S$-wave scattering lengths~\cite{Ananthanarayan:2000ht}. In the case of $\pi N$ scattering, the construction and solution is complicated by the presence of the crossed channel $\pi\pi\to N \bar N$, cf.~\cite{Hite:1973pm,Buettiker:2003pp,Hoferichter:2011wk}, as well as the increased number of relevant partial waves. While partial results have already been presented in~\cite{Ditsche:2012fv,Ditsche:2012ja,Elvira:2014wma,Elvira:2014lta}, here we use the complete solution of the RS system to obtain, in combination with the scattering-length constraints from pionic atoms, a precision determination of the $\pi N$ $\sigma$-term. In particular, at this level of accuracy the impact of isospin-violating (IV) corrections cannot be ignored, as demonstrated by the isoscalar $\pi N$ scattering length~\cite{Baru:2010xn,Baru:2011bw,Gasser:2002am,Hoferichter:2009ez}, so that revisiting the Cheng--Dashen LET becomes mandatory.       

\section{Cheng--Dashen low-energy theorem}

We start by stating the precise formulation of the Cheng--Dashen LET~\cite{Cheng:1970mx,Brown:1971pn}. In the isospin limit,
the scattering amplitude for the process
\beq
\pi^a(q)+N(p)\to\pi^b(q')+N(p'),
\eeq
with pion isospin labels $a$, $b$ and Mandelstam variables
\beq
s=(p+q)^2,\qquad t=(p'-p)^2,\qquad u=(p-q')^2,
\eeq
fulfilling $s+t+u=2\mN^2+2\mpi^2$,
can be expressed as
\begin{align}
T^{ba}(\nu,t)&=\delta^{ba}T^+(\nu,t)+\frac{1}{2}[\tau^b,\tau^a]T^-(\nu,t),\notag\\
T^I(\nu,t)&=\bar{u}(p')\bigg\{D^I(\nu,t)-\frac{[\slashed q',\slashed q]}{4\mN}B^I(\nu,t)\bigg\}u(p),
\end{align}
where  $\nu=(s-u)/(4\mN)$, $I=\pm$ refers to isoscalar/isovector amplitudes, $\mN$ and $\mpi$ to the nucleon and pion mass, $\tau^a$ denotes isospin Pauli matrices, and we normalize spinors as $\bar u u=1$. Amplitudes $\Amp^{I_s}$ with definite $s$-channel isospin $I_s$ are
\beq
\begin{pmatrix}\Amp^{1/2}\\\Amp^{3/2}\end{pmatrix}=\begin{pmatrix}1&2\\1&-1\end{pmatrix}\begin{pmatrix}\Amp^+\\\Amp^-\end{pmatrix},\qquad
\Amp\in\{D,B\}.
\eeq
The LET involves the Born-term-subtracted amplitude
\beq
\bar D^+(\nu,t)=D^+(\nu,t)-\frac{g^2}{\mN}-\nu g^2\bigg(\frac{1}{\mN^2-s}-\frac{1}{\mN^2-u}\bigg),
\eeq
where $g$ is the $\pi N$ coupling constant. 
The scalar form factor of the nucleon is defined as the matrix element
\beq
\sigma(t)=\langle N(p')|\hat m(\bar u u+\bar d d)|N(p)\rangle,\qquad \hat m=\frac{\muu+\md}{2},
\eeq
with up- and down-quark masses $\muu$ and $\md$, momentum transfer $t=(p'-p)^2$, and $\sigma(0)=\sigma_{\pi N}$. The LET then states that 
\beq
\label{LET}
\bar D^+(0,2\mpi^2)=\sigma(2\mpi^2)+\Delta_R,
\eeq
where $\Delta_R$ subsumes higher-order corrections in the chiral expansion. Corrections to the LET have been investigated systematically in $SU(2)$ ChPT, with the result that $\Delta_R$ is very small: non-analytic terms are absent at full one-loop order~\cite{Bernard:1996nu,Becher:2001hv}, so that the dominant corrections are expected to scale as $\mpi^2/\mN^2 \sigma_{\pi N}\sim 1\MeV$. Indeed, estimating the low-energy constants based on resonance exchange, one obtains~\cite{Bernard:1996nu}
\beq
|\Delta_R|\lesssim 2\MeV,
\eeq
an estimate which we will adopt in the following. 

In practice, \eqref{LET} is usually rewritten as
\beq
\sigma_{\pi N}=\Sigma_d+\Delta_D-\Delta_\sigma-\Delta_R,
\eeq
where 
\begin{align}
 \Delta_\sigma&=\sigma(2\mpi^2)-\sigma_{\pi N},\qquad \Delta_D=\bar D^+(0,2\mpi^2)-\Sigma_d,\notag\\
\Sigma_d&=\Fpi^2\big(d_{00}^++2\mpi^2d_{01}^+\big).
\end{align}
Here, $\Fpi=92.2\MeV$~\cite{PDG} denotes the pion decay constant and the subthreshold coefficients are defined via the expansion
\beq
\bar D^+(\nu,t)=\sum_{n,m=0}^\infty d_{mn}^+\nu^{2m}t^n.
\eeq
Although individually sizable due to strong $\pi\pi$ rescattering, the difference $\Delta_D-\Delta_\sigma$ was shown to be small in~\cite{Gasser:1990ap}. Here, we use the updated value~\cite{Hoferichter:2012wf,Hoferichter:2012tu}
\beq
\Delta_D-\Delta_\sigma=(-1.8\pm 0.2)\MeV,
\eeq
which incorporates modern input for $\pi\pi$ phase shifts, effects from $K\bar K$ intermediate states, and the uncertainties due to $\pi N$ parameters.

As alluded to above, the isoscalar channel is known to be sensitive to IV corrections. For this reason, we now derive a version of the LET that takes the dominant IV effects into account. First, we define the $\sigma$-term as the average value of proton and neutron scalar-current matrix elements ($N\in\{p,n\}$)
\beq
\sigma_{\pi N}=\frac{\sigma_p+\sigma_n}{2},\qquad \sigma_N=\langle N|\hat m(\bar u u+\bar d d)|N\rangle,
\eeq
where, up to third order in the chiral expansion, one finds $\sigma_p=\sigma_n$~\cite{Meissner:1997ii}. Next, we identify the isoscalar amplitudes everywhere with the average of the $\pi^\pm p\to\pi^\pm p$ charge channels
\beq
X^+\to X^p=\frac{1}{2}\big(X_{\pi^+p\to\pi^+ p}+X_{\pi^-p\to\pi^- p}\big),
\eeq
for $X\in\{D,d_{00},d_{01},\ldots\}$. The motivation for doing so is two-fold: first, the $\pi^\pm p$ charge channels dominate the $\pi N$ data base, so that this scenario is closest to the one considered in PWAs. Second, the uncertainties in the $\pi N$ scattering lengths are smallest if one works in the physical, not the isospin basis~\cite{Baru:2010xn,Baru:2011bw}. As a consequence, we identify the nucleon and pion mass with the masses of the proton and the charged pion, respectively. We also assume that the radiative corrections applied in the PWAs remove the dominant effects, and therefore we consider all quantities to be virtual-photon subtracted. In this scenario, the leading IV corrections are generated by the mass difference between charged and neutral pion $\Delta_\pi=\mpi^2-\mpii^2$. For the scalar form factor one finds~\cite{Meissner:1997ii}
\begin{align}
\Delta_{\sigma}^p&=\sigma_p\big(2\mpi^2\big)-\sigma_p\\
&=\frac{3\ga^2\mpi^3}{64\pi\Fpi^2}+\frac{\ga^2\mpi\Delta_\pi}{128\pi\Fpi^2}\Big(-7+2\sqrt{2}\log\big(1+\sqrt{2}\big)\Big),\notag
\end{align}
where $\ga$ denotes the axial charge of the nucleon. Similarly, the IV corrections to $\Delta_D$ can be extracted from~\cite{Hoferichter:2009gn}
\begin{align}
 \Delta_D^p&=\Fpi^2\Big\{D_p\big(0,2\mpi^2\big)-d_{00}^p-2\mpi^2d_{01}^p\Big\}\\
 &=\frac{23\ga^2\mpi^3}{384\pi\Fpi^2}+\frac{\ga^2\mpi\Delta_\pi}{256\pi\Fpi^2}\Big(3+4\sqrt{2}\log\big(1+\sqrt{2}\big)\Big).\notag
\end{align}
Taking everything together, we obtain
\begin{align}
\label{LET_IV_corr}
\sigma_{\pi N}&=\Fpi^2\big(d_{00}^p+2\mpi^2d_{01}^p\big)+\Delta_D-\Delta_\sigma-\Delta_R\notag\\
 &+\frac{81\ga^2\mpi\Delta_\pi}{256\pi\Fpi^2}+\frac{e^2}{2}\Fpi^2\big(4f_1+f_2\big)\notag\\
 &=\Fpi^2\big(d_{00}^p+2\mpi^2d_{01}^p\big)+(1.2\pm 3.0)\MeV.
\end{align}
In~\eqref{LET_IV_corr} we also included the leading corrections due to virtual photons, encoded in the low-energy constants $f_1$ and $f_2$. The latter can be determined from the proton--neutron mass difference~\cite{Gasser:1974wd}, $f_2=(-0.97\pm 0.38)\GeV^{-1}$, for the former we use the estimate $|f_1|\leq 1.4\GeV^{-1}$~\cite{Fettes:2000vm,Gasser:2002am}. The single largest correction is generated by $\Delta_\pi$, an upward shift of $3.4\MeV$. Such large IV corrections have already been observed in the case of the $\pi N$ scattering lengths~\cite{Gasser:2002am,Hoferichter:2009ez}.

\section{Pionic atoms}

Pionic hydrogen ($\pi H$) and deuterium ($\pi D$), a $\pi^-$ and a proton/deuteron bound by electromagnetism, provide access to $\pi N$ physics due to the imprint of strong interactions in the energy spectrum.
The shift of the ground-state energy level in $\pi H$ and $\pi D$, as well as the width of the $\pi H$ ground state, probe three different combinations of $\pi N$ scattering lengths. The input quantities relevant for the RS equations are the $s$-channel-isospin scattering lengths $a^{I_s}_{0+}$, defined in terms of the $\pi^\pm p$ charge channels. Updating the analysis of~\cite{Baru:2010xn,Baru:2011bw} to account for the new value of the $\pi H$ level shift~\cite{Hennebach:2014lsa} and subtracting virtual-photon effects as detailed in~\cite{Hoferichter:2012bz}, we obtain
\begin{align}
\label{scatt_length}
a^{1/2}_{0+}&=(169.8\pm 2.0)\times 10^{-3}\mpi^{-1},\notag\\ 
a^{3/2}_{0+}&=(-86.3\pm 1.8)\times 10^{-3}\mpi^{-1}.
\end{align}
Since the errors are dominated by different sources---IV corrections in the case of $a^{1/2}_{0+}$ and uncertainty in the extraction of the isoscalar combination for $a^{3/2}_{0+}$---the errors can be considered uncorrelated to a very good approximation.
Apart from their role in the solution of the RS equations,
the scattering lengths are also a crucial ingredient in the determination of the $\pi N$ coupling constant via the Goldberger--Miyazawa--Oehme sum rule~\cite{Goldberger:1955zza}. Indeed, if the scattering lengths from~\cite{Koch:1980ay,Hoehler} are used, one recovers the value $g^2/(4\pi)=14.3$, whereas~\eqref{scatt_length} leads to $g^2/(4\pi)=13.7\pm 0.2$~\cite{Baru:2010xn,Baru:2011bw}. This result, to be adopted in the following, stands in good agreement with more recent determinations from $NN$~\cite{deSwart:1997ep} and $\pi N$~\cite{Arndt:2006bf} scattering.

\section{Roy--Steiner equations}

Roy equations~\cite{Roy:1971tc} for $\pi\pi$ scattering, or RS equations~\cite{Hite:1973pm,Buettiker:2003pp,Hoferichter:2011wk,Ditsche:2012fv} for non-totally-crossing-symmetric processes, incorporate the constraints from analyticity, unitarity, and crossing symmetry in the form of dispersion relations for the partial waves. They can be shown to be rigorously valid in a certain kinematic region, in the case of $\pi N$ scattering the upper limit is $\sm=(1.38\GeV)^2$~\cite{Ditsche:2012fv}. The integral contributions above $\sm$ as well as partial waves with $l>l_\text{m}$, with $l_\text{m}$ the maximal angular momentum explicitly included in the calculation, are collected in the so-called driving terms, which need to be estimated from existing PWAs, as do inelastic contributions below $\sm$. The free parameters of the approach are subtraction constants, which, in the case of $\pi\pi$ scattering, can be directly identified with the scattering lengths~\cite{Ananthanarayan:2000ht}, while for the solution of the $\pi N$ system it is more convenient to relate them to subthreshold parameters instead. The resulting system of coupled integral equations corresponds to a self-consistency condition for the low-energy phase shifts, whose mathematical properties were investigated in detail in~\cite{Gasser:1999hz}. Following~\cite{Ananthanarayan:2000ht}, we pursue the following solution strategy: the phase shifts are parameterized in a convenient way with a few parameters each, which are matched to input partial waves above $\sm$ in a smooth way.
To measure the degree to which the RS are fulfilled, a $\chi^2$-like function is defined according to
\beq
\label{chisqr}
\chi^2=\sum_{l,I_s,\pm}\sum_{j=1}^N\Bigg(\frac{\Re f_{l\pm}^{I_s}(W_j)-F\big[f_{l\pm}^{I_s}\big](W_j)}{\Re f_{l\pm}^{I_s}(W_j)}\Bigg)^2,
\eeq
where $\{W_j\}$ denotes a set of points between threshold and $\sqrt{\sm}$, $f_{l\pm}^{I_s}$ are the $s$-channel partial waves with isospin $I_s$, orbital angular momentum $l$, and total angular momentum $j=l\pm 1/2\equiv l\pm$, and $F\big[f_{l\pm}^{I_s}\big]$ the right-hand side of the RS equations.
We take $l_\text{m}=1$, $N=25$ (distributed equidistantly), and choose the number of subtraction constants in such a way as to match the number of degrees of freedom predicted by the mathematical properties of the Roy equations~\cite{Gasser:1999hz}. It should be stressed that the form of the RS equations only reduces to that of Roy equations once the $t$-channel is solved, see~\cite{Ditsche:2012fv}. In the solution of the RS equations we minimize~\eqref{chisqr} with respect to the subtraction constants (identified with subthreshold parameters) and the parameters describing the low-energy phase shifts, while imposing~\eqref{scatt_length} as additional constraints.

\begin{figure}
\centering
\includegraphics[height=\linewidth,angle=-90]{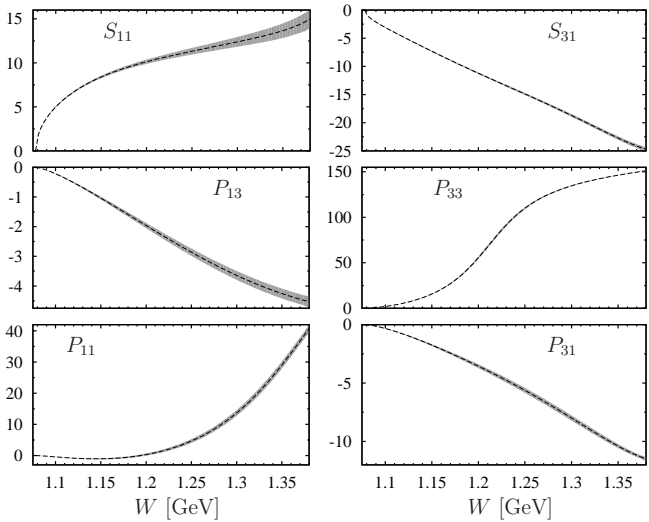}
\caption{Phase shifts $\delta_{l\pm}^{I_s}$ of the $s$-channel partial waves in degrees, obtained from the solution of the RS equations. 
The dashed line indicates our central solution, the bands the uncertainty estimate.
The partial waves are labeled by the spectroscopic notation $L_{2I_s2J}$.}
\label{fig:schannel}
\end{figure}

We performed a number of checks as regards the sensitivity of our solution to the input quantities: the number of grid points $N$ as well as the number of parameters used in the description of the partial waves were varied, the $s$- and $t$-channel partial waves in the driving terms truncated at different $l_\text{max}=4,5$ and $J_\text{max}=2,3$, the matching conditions at $\sm$ as well as the $s$-channel partial waves evaluated from different PWAs, and the sensitivity to the precise definition of the $\chi^2$-function was investigated. In addition, to stabilize the fit we imposed sum rules for the higher subthreshold parameters.   
The solution for the $s$-channel partial waves, expressed in terms of the phase shifts and including uncertainty estimates from these systematic studies as well as the uncertainties in the scattering lengths and the coupling constant, is shown in Fig.~\ref{fig:schannel}. A more detailed account of our RS solution will be given in~\cite{inprep}. 

Apart from low-energy phase shifts, the RS solution provides a consistent set of subthreshold parameters. In particular, this allows us to pin down $\Sigma_d$ in accord with both the RS and the scattering-length constraints. Linearizing around the central values~\eqref{scatt_length}, we find
\begin{align}
\label{Sigma_d_lin}
\Sigma_d&=(57.9\pm 0.9)\MeV + \sum_{I_s}c_{I_s}\Delta a^{I_s}_{0+},\notag\\
c_{1/2}&=0.24\MeV,\qquad c_{3/2}=0.89\MeV,
\end{align}
where $\Delta a^{I_s}_{0+}$ measures the deviation from~\eqref{scatt_length} in units of $10^{-3}\mpi^{-1}$.
Already in this linearized form, one recovers $\Sigma_d$ from~\cite{Gasser:1990ce} if the scattering lengths from~\cite{Koch:1980ay,Hoehler} are inserted, while the modern input produces $\Sigma_d=(57.9\pm 1.9)\MeV$ (this also indicates that the $S$-wave phase shifts from~\cite{Koch:1980ay,Hoehler} need to be amended close to threshold).
Moreover, the difference to~\cite{Pavan:2001wz} can be traced back to the $P$-wave scattering volume $a_{1+}^+$, which needs to be known extremely accurately due to its large weight in the formalism of~\cite{Gasser:1988jt,Gasser:1990ce}. Once the RS equations are solved, the threshold parameters can be calculated from sum rules, and indeed we find that the result for $a_{1+}^+$ is slightly lower than the value used in~\cite{Pavan:2001wz}, which already suffices to explain the difference~\cite{inprep}. The main impact of the RS equations in the $\sigma$-term determination thus amounts to eliminating the need for independent input for $a_{1+}^+$.
In total, our result for the $\sigma$-term becomes
\beq
\sigma_{\pi N}=(59.1\pm 3.5)\MeV.
\eeq
Although already $4.2\MeV$ are due to new corrections to the LET (thereof $3.0\MeV$ from isospin breaking), we do observe a significant increase
compared to~\cite{Gasser:1990ce}. As illustrated by~\eqref{Sigma_d_lin}, this effect can be immediately traced back to 
our modern knowledge of the $\pi N$ scattering lengths as extracted from pionic atoms. By combining this information
with the constraints from RS equations, the $\sigma$-term can be determined to a remarkable accuracy.

\section{Scalar nucleon couplings}

The existence of a weakly-interacting massive particle (WIMP), one of the most promising dark-matter candidates, could be established in direct-detection experiments,
which are sensitive to the recoil of the WIMP scattering off nuclei (see~\cite{Cushman:2013zza} for a review).
The interpretation of these searches relies on the couplings of the WIMP to nucleons, according to its quantum numbers. 
A precise determination of $\sigma_{\pi N}$ therefore has immediate consequences for the scalar channel, since,
as it was shown in~\cite{Crivellin:2013ipa},
the scalar couplings of the nucleon to $q=u,d$,
\beq
\mN f_q^N=\langle N|m_q\bar q q|N\rangle,
\eeq
follow once $\sigma_{\pi N}$ is determined, with all further corrections taken into account within $SU(2)$ ChPT.
Taking $\muu/\md= 0.46\pm0.03$ from~\cite{Aoki:2013ldr}, we obtain
\begin{align}
\label{fs}
 f_u^p&=(20.8\pm 1.5)\times 10^{-3},\qquad  f_d^p=(41.1\pm 2.8)\times 10^{-3},\notag\\
 f_u^n&=(18.9\pm 1.4)\times 10^{-3},\qquad  f_d^n=(45.1\pm 2.7)\times 10^{-3}.
\end{align}
In addition, we quote our result for
\beq
\sum_{q=u,\ldots,t} f_q^N=\frac{2}{9}+\frac{7}{9}\big(f_u^N+f_d^N+f_s^N\big)=0.305\pm 0.009,
\eeq
averaged over proton and neutron, and with $f_s^N$ taken from~\cite{Junnarkar:2013ac} (in principle, the strangeness coupling follows from the $\sigma$-term by means of $SU(3)$ considerations, but the uncertainties are too large to compete with recent lattice determinations). This 
particular combination of scalar coefficients becomes relevant in the context of
Higgs-mediated interactions, not only in direct detection, but also in Higgs-induced
lepton flavor violation~\cite{Crivellin:2014cta}.
In particular in~\eqref{fs} the uncertainties have been appreciably reduced, thanks to 
the precise knowledge of $\sigma_{\pi N}$ inferred from our RS equation analysis of $\pi N$ scattering.

\section*{Acknowledgments}

We thank Christoph Ditsche for collaboration at early stages of this project,
and Gilberto Colangelo for many helpful discussions.
Financial support by
BMBF ARCHES, the Helmholtz Alliance HA216/EMMI, 
 the Helmholtz Virtual Institute NAVI (VH-VI-417),
the Swiss National Science Foundation,
the DFG (SFB/TR 16, ``Subnuclear Structure of Matter''),
EU I3HP ``Study of Strongly Interacting Matter,'' and 
the DOE (Grant No.\ DE-FG02-00ER41132) 
is gratefully acknowledged.


\begin{thebibliography}{99}

\bibitem{Bottino:1999ei} 
  A.~Bottino, F.~Donato, N.~Fornengo and S.~Scopel,
  %``Implications for relic neutralinos of the theoretical uncertainties in the neutralino nucleon cross-section,''
  Astropart.\ Phys.\  {\bf 13}, 215 (2000)
  [hep-ph/9909228].
  %%CITATION = HEP-PH/9909228;%%
  
\bibitem{Ellis:2008hf} 
  J.~R.~Ellis, K.~A.~Olive and C.~Savage,
  %``Hadronic Uncertainties in the Elastic Scattering of Supersymmetric Dark Matter,''
  Phys.\ Rev.\ D {\bf 77}, 065026 (2008)
  [arXiv:0801.3656 [hep-ph]].
  %%CITATION = ARXIV:0801.3656;%%  

\bibitem{Crivellin:2013ipa} 
  A.~Crivellin, M.~Hoferichter and M.~Procura,
  %``Accurate evaluation of hadronic uncertainties in spin-independent WIMP-nucleon scattering: Disentangling two- and three-flavor effects,''
  Phys.\ Rev.\ D {\bf 89}, 054021 (2014)
  [arXiv:1312.4951 [hep-ph]].
  %%CITATION = ARXIV:1312.4951;%%

\bibitem{Cheng:1970mx} 
  T.~P.~Cheng and R.~F.~Dashen,
  %``Is SU(2) x SU(2) a better symmetry than SU(3)?,''
  Phys.\ Rev.\ Lett.\  {\bf 26}, 594 (1971).
  %%CITATION = PRLTA,26,594;%%
  
\bibitem{Brown:1971pn} 
  L.~S.~Brown, W.~J.~Pardee and R.~D.~Peccei,
  %``Adler-Weisberger theorem reexamined,''
  Phys.\ Rev.\ D {\bf 4}, 2801 (1971).
  %%CITATION = PHRVA,D4,2801;%%
  
\bibitem{Koch:1980ay} 
  R.~Koch and E.~Pietarinen,
  %``Low-Energy pi N Partial Wave Analysis,''
  Nucl.\ Phys.\ A {\bf 336}, 331 (1980).
  %%CITATION = NUPHA,A336,331;%% 
  
\bibitem{Koch:1982pu} 
  R.~Koch,
  %``A New Determination of the pi N Sigma Term Using Hyperbolic Dispersion Relations in the (nu**2, t) Plane,''
  Z.\ Phys.\ C {\bf 15}, 161 (1982).
  %%CITATION = ZEPYA,C15,161;%%   
  
\bibitem{Hoehler}
  G.~H\"{o}hler,
  {\it Pion--Nukleon-Streuung: Methoden und Ergebnisse},
  in Landolt-B\"ornstein, {\bf 9b2}, ed.\ H.~Schopper,
  Springer Verlag, Berlin, 1983.  
  
\bibitem{Gasser:1988jt} 
  J.~Gasser, H.~Leutwyler, M.~P.~Locher and M.~E.~Sainio,
  %``Extracting the Pion - Nucleon $\Sigma$ Term From Data,''
  Phys.\ Lett.\ B {\bf 213}, 85 (1988).
  %%CITATION = PHLTA,B213,85;%%  
  
\bibitem{Gasser:1990ce} 
  J.~Gasser, H.~Leutwyler and M.~E.~Sainio,
  %``Sigma term update,''
  Phys.\ Lett.\ B {\bf 253}, 252 (1991).
  %%CITATION = PHLTA,B253,252;%%
  
\bibitem{Gasser:1990ap} 
  J.~Gasser, H.~Leutwyler and M.~E.~Sainio,
  %``Form-factor of the sigma term,''
  Phys.\ Lett.\ B {\bf 253}, 260 (1991).
  %%CITATION = PHLTA,B253,260;%%  
  
\bibitem{Pavan:2001wz} 
  M.~M.~Pavan, I.~I.~Strakovsky, R.~L.~Workman and R.~A.~Arndt,
  %``The Pion nucleon Sigma term is definitely large: Results from a G.W.U. analysis of pi nucleon scattering data,''
  PiN Newslett.\  {\bf 16}, 110 (2002)
  [hep-ph/0111066].
  %%CITATION = HEP-PH/0111066;%%
  
\bibitem{Fettes:2000xg} 
  N.~Fettes and U.-G.~Mei\ss ner,
  %``Pion nucleon scattering in chiral perturbation theory. 2.: Fourth order calculation,''
  Nucl.\ Phys.\ A {\bf 676}, 311 (2000)
  [hep-ph/0002162].
  %%CITATION = HEP-PH/0002162;%%
  
\bibitem{Alarcon:2011zs} 
  J.~M.~Alarc\'on, J.~Martin Camalich and J.~A.~Oller,
  %``The chiral representation of the $\pi N$ scattering amplitude and the pion-nucleon sigma term,''
  Phys.\ Rev.\ D {\bf 85}, 051503 (2012)
  [arXiv:1110.3797 [hep-ph]].
  %%CITATION = ARXIV:1110.3797;%%
  
\bibitem{Becher:2001hv} 
  T.~Becher and H.~Leutwyler,
  %``Low energy analysis of pi N ---> pi N,''
  JHEP {\bf 0106}, 017 (2001)
  [hep-ph/0103263].
  %%CITATION = HEP-PH/0103263;%%
  
\bibitem{inprep}
 M.~Hoferichter, J.~Ruiz de Elvira, B.~Kubis and U.-G.~Mei{\ss}ner,
 in preparation.  
  
\bibitem{Gotta:2008zza} 
  D.~Gotta {\it et al.},
  %``Pionic hydrogen,''
  Lect.\ Notes Phys.\  {\bf 745}, 165 (2008).
  %%CITATION = LNPHA,745,165;%%  
  
\bibitem{Strauch:2010vu} 
  T.~Strauch {\it et al.},
  %``Pionic deuterium,''
  Eur.\ Phys.\ J.\ A {\bf 47}, 88 (2011)
  [arXiv:1011.2415 [nucl-ex]].
  %%CITATION = ARXIV:1011.2415;%%  
  
\bibitem{Hennebach:2014lsa} 
  M.~Hennebach {\it et al.},
  %``Hadronic shift in pionic hydrogen,''
  Eur.\ Phys.\ J.\ A {\bf 50}, 190 (2014)
  [arXiv:1406.6525 [nucl-ex]].
  %%CITATION = ARXIV:1406.6525;%%
  
\bibitem{Baru:2010xn} 
  V.~Baru {\it et al.},
  %``Precision calculation of the pi^- deuteron scattering length and its impact on threshold pi-N scattering,''
  Phys.\ Lett.\ B {\bf 694}, 473 (2011)
  [arXiv:1003.4444 [nucl-th]].
  %%CITATION = ARXIV:1003.4444;%%  
  
\bibitem{Baru:2011bw} 
  V.~Baru {\it et al.},
  %``Precision calculation of threshold pi^- d scattering, pi N scattering lengths, and the GMO sum rule,''
  Nucl.\ Phys.\ A {\bf 872}, 69 (2011)
  [arXiv:1107.5509 [nucl-th]].
  %%CITATION = ARXIV:1107.5509;%%  

 \bibitem{Ditsche:2012fv} 
  C.~Ditsche, M.~Hoferichter, B.~Kubis and U.-G.~Mei{\ss}ner,
  %``Roy-Steiner equations for pion-nucleon scattering,''
  JHEP {\bf 1206}, 043 (2012)
  [arXiv:1203.4758 [hep-ph]].
  %%CITATION = ARXIV:1203.4758;%%
  
\bibitem{Roy:1971tc} 
  S.~M.~Roy,
  %``Exact integral equation for pion pion scattering involving only physical region partial waves,''
  Phys.\ Lett.\ B {\bf 36}, 353 (1971).
  %%CITATION = PHLTA,B36,353;%%
  
\bibitem{Ananthanarayan:2000ht} 
  B.~Ananthanarayan, G.~Colangelo, J.~Gasser and H.~Leutwyler,
  %``Roy equation analysis of pi pi scattering,''
  Phys.\ Rept.\  {\bf 353}, 207 (2001)
  [hep-ph/0005297].
  %%CITATION = HEP-PH/0005297;%%

\bibitem{Hite:1973pm} 
  G.~E.~Hite and F.~Steiner,
  %``New dispersion relations and their application to partial-wave amplitudes,''
  Nuovo Cim.\ A {\bf 18}, 237 (1973).
  %%CITATION = NUCIA,A18,237;%%  
  
\bibitem{Buettiker:2003pp} 
  P.~B\"uttiker, S.~Descotes-Genon and B.~Moussallam,
  %``A new analysis of pi K scattering from Roy and Steiner type equations,''
  Eur.\ Phys.\ J.\ C {\bf 33}, 409 (2004)
  [hep-ph/0310283].
  %%CITATION = HEP-PH/0310283;%%  
  
\bibitem{Hoferichter:2011wk} 
  M.~Hoferichter, D.~R.~Phillips and C.~Schat,
  %``Roy-Steiner equations for gamma gamma -> pi pi,''
  Eur.\ Phys.\ J.\ C {\bf 71}, 1743 (2011)
  [arXiv:1106.4147 [hep-ph]].
  %%CITATION = ARXIV:1106.4147;%%  

\bibitem{Ditsche:2012ja} 
  C.~Ditsche, M.~Hoferichter, B.~Kubis and U.-G.~Mei{\ss}ner,
  %``Roy–Steiner equations for π N scattering,''
  PoS {\bf CD 12}, 064 (2013)
  [arXiv:1211.7285 [hep-ph]].
  %%CITATION = ARXIV:1211.7285;%%  
  
\bibitem{Elvira:2014wma} 
  J.~Ruiz de Elvira, C.~Ditsche, M.~Hoferichter, B.~Kubis and U.-G.~Mei{\ss}ner,
  %``Roy-Steiner equations for piN scattering,''
  EPJ Web Conf.\  {\bf 73}, 05002 (2014).
  %%CITATION = 00776,73,05002;%% 
  
\bibitem{Elvira:2014lta} 
  J.~Ruiz de Elvira, C.~Ditsche, M.~Hoferichter, B.~Kubis and U.-G.~Mei{\ss}ner,
  %``Roy–Steiner equations for πN scattering,''
  %``Seventh International Symposium on Chiral Symmetry in Hadrons and Nuclei 2013. Proceedings, Conference Beijing, China, October 27-30, 2013,''
  Singapore, Singapore: World Scientific (2014) 186.
  %%CITATION = INSPIRE-1330197;%%
  
\bibitem{Gasser:2002am} 
  J.~Gasser, M.~A.~Ivanov, E.~Lipartia, M.~Moj\v zi\v s and A.~Rusetsky,
  %``Ground state energy of pionic hydrogen to one loop,''
  Eur.\ Phys.\ J.\ C {\bf 26}, 13 (2002)
  [hep-ph/0206068].
  %%CITATION = HEP-PH/0206068;%%
  
\bibitem{Hoferichter:2009ez} 
  M.~Hoferichter, B.~Kubis and U.-G.~Mei{\ss}ner,
  %``Isospin breaking in the pion-nucleon scattering lengths,''
  Phys.\ Lett.\ B {\bf 678}, 65 (2009)
  [arXiv:0903.3890 [hep-ph]].
  %%CITATION = ARXIV:0903.3890;%%
  
\bibitem{Bernard:1996nu} 
  V.~Bernard, N.~Kaiser and U.-G.~Mei{\ss}ner,
  %``On the analysis of the pion - nucleon sigma term: The Size of the remainder at the Cheng-Dashen point,''
  Phys.\ Lett.\ B {\bf 389}, 144 (1996)
  [hep-ph/9607245].
  %%CITATION = HEP-PH/9607245;%%  
  
\bibitem{PDG} 
  K.~A.~Olive {\it et al.}  [Particle Data Group Collaboration],
  %``Review of Particle Physics,''
  Chin.\ Phys.\ C {\bf 38}, 090001 (2014).
  %%CITATION = CHPHD,C38,090001;%%
  
\bibitem{Hoferichter:2012wf} 
  M.~Hoferichter, C.~Ditsche, B.~Kubis and U.-G.~Mei{\ss}ner,
  %``Dispersive analysis of the scalar form factor of the nucleon,''
  JHEP {\bf 1206}, 063 (2012)
  [arXiv:1204.6251 [hep-ph]].
  %%CITATION = ARXIV:1204.6251;%% 
  
\bibitem{Hoferichter:2012tu} 
  M.~Hoferichter, C.~Ditsche, B.~Kubis and U.-G.~Mei{\ss}ner,
  %``Improved dispersive analysis of the scalar form factor of the nucleon,''
  PoS {\bf CD  12}, 069 (2013)
  [arXiv:1211.1485 [nucl-th]].
  %%CITATION = ARXIV:1211.1485;%%
  
\bibitem{Meissner:1997ii} 
  U.-G.~Mei{\ss}ner and S.~Steininger,
  %``Isospin violation in pion nucleon scattering,''
  Phys.\ Lett.\ B {\bf 419}, 403 (1998)
  [hep-ph/9709453].
  %%CITATION = HEP-PH/9709453;%%
  
\bibitem{Hoferichter:2009gn} 
  M.~Hoferichter, B.~Kubis and U.-G.~Mei{\ss}ner,
  %``Isospin Violation in Low-Energy Pion-Nucleon Scattering Revisited,''
  Nucl.\ Phys.\ A {\bf 833}, 18 (2010)
  [arXiv:0909.4390 [hep-ph]].
  %%CITATION = ARXIV:0909.4390;%%
  
\bibitem{Gasser:1974wd} 
  J.~Gasser and H.~Leutwyler,
  %``Implications of Scaling for the Proton - Neutron Mass - Difference,''
  Nucl.\ Phys.\ B {\bf 94}, 269 (1975).
  %%CITATION = NUPHA,B94,269;%%
  
\bibitem{Fettes:2000vm} 
  N.~Fettes and U.-G.~Mei{\ss}ner,
  %``Towards an understanding of isospin violation in pion nucleon scattering,''
  Phys.\ Rev.\ C {\bf 63}, 045201 (2001)
  [hep-ph/0008181].
  %%CITATION = HEP-PH/0008181;%%
  
\bibitem{Hoferichter:2012bz} 
  M.~Hoferichter {\it et al.},
  %``Isospin breaking in pion–deuteron scattering and the pion–nucleon scattering lengths,''
  PoS {\bf CD  12}, 093 (2013)
  [arXiv:1211.1145 [nucl-th]].
  %%CITATION = ARXIV:1211.1145;%%
  
\bibitem{Goldberger:1955zza} 
  M.~L.~Goldberger, H.~Miyazawa and R.~Oehme,
  %``Application of Dispersion Relations to Pion-Nucleon Scattering,''
  Phys.\ Rev.\  {\bf 99}, 986 (1955).
  %%CITATION = PHRVA,99,986;%%
  
\bibitem{deSwart:1997ep} 
  J.~J.~de Swart, M.~C.~M.~Rentmeester and R.~G.~E.~Timmermans,
  %``The Status of the pion - nucleon coupling constant,''
  PiN Newslett.\  {\bf 13}, 96 (1997)
  [nucl-th/9802084].
  %%CITATION = NUCL-TH/9802084;%%
  
\bibitem{Arndt:2006bf} 
  R.~A.~Arndt, W.~J.~Briscoe, I.~I.~Strakovsky and R.~L.~Workman,
  %``Extended partial-wave analysis of piN scattering data,''
  Phys.\ Rev.\ C {\bf 74}, 045205 (2006)
  [nucl-th/0605082].
  %%CITATION = NUCL-TH/0605082;%%
  
\bibitem{Gasser:1999hz} 
  J.~Gasser and G.~Wanders,
  %``One channel Roy equations revisited,''
  Eur.\ Phys.\ J.\ C {\bf 10}, 159 (1999)
  [hep-ph/9903443].
  %%CITATION = HEP-PH/9903443;%%
  
\bibitem{Cushman:2013zza} 
  P.~Cushman {\it et al.},
  %``Working Group Report: WIMP Dark Matter Direct Detection,''
  arXiv:1310.8327 [hep-ex].
  %%CITATION = ARXIV:1310.8327;%%
 
\bibitem{Aoki:2013ldr} 
  S.~Aoki {\it et al.},
  %``Review of lattice results concerning low-energy particle physics,''
  Eur.\ Phys.\ J.\ C {\bf 74}, 2890 (2014)
  [arXiv:1310.8555 [hep-lat]].
  %%CITATION = ARXIV:1310.8555;%% 
  
\bibitem{Junnarkar:2013ac} 
  P.~Junnarkar and A.~Walker-Loud,
  %``Scalar strange content of the nucleon from lattice QCD,''
  Phys.\ Rev.\ D {\bf 87}, 114510 (2013)
  [arXiv:1301.1114 [hep-lat]].
  %%CITATION = ARXIV:1301.1114;%%
  
\bibitem{Crivellin:2014cta} 
  A.~Crivellin, M.~Hoferichter and M.~Procura,
  %``Improved predictions for $\mu\to e$ conversion in nuclei and Higgs-induced lepton flavor violation,''
  Phys.\ Rev.\ D {\bf 89}, 093024 (2014)
  [arXiv:1404.7134 [hep-ph]].
  %%CITATION = ARXIV:1404.7134;%% 
  
\end{thebibliography}
\end{document}